\documentclass{article}
\usepackage{spconf,amsmath,graphicx}
\usepackage{subcaption,subfiles}
\usepackage{svg}
\usepackage{url,xcolor}
\usepackage{hyperref}
\usepackage{fancyhdr}

\title{Semantics-Guided Generative Image Compression}

\name{Cheng-Lin Wu,$^{\ast}$ Hyomin Choi,$^{\dag}$ and Ivan V. Baji\'{c}$^{\ast}$}
\address{$^{\ast}$Simon Fraser University, Burnaby, BC, Canada \hspace{1cm} $^{\dag}$InterDigital, Los Altos, CA, USA\\
}

  \fancypagestyle{firstpage}{
    \renewcommand{\topmargin}{-13mm}
    \setlength{\headsep}{2.0\normalbaselineskip}

    \renewcommand{\headsep}{15mm}
    \fancyhead{}
    \fancyfoot{}
    \fancyhead[C]{%
      \footnotesize%
      To be presented at \textit{IEEE ICIP 2025}, Anchorage, AK, USA, Sept. 2025.\\
      \vspace{0.6\normalbaselineskip}%
      \tiny%
      \copyright{} Copyright 2025 IEEE. Published in 2025 IEEE International Conference on Image Processing (ICIP), scheduled for 14-17 September 2025 in Anchorage, Alaska, USA. Personal use of this material is permitted. However, permission to reprint/republish this material for advertising or promotional purposes or for creating new collective works for resale or redistribution to servers or lists, or to reuse any copyrighted component of this work in other works, must be obtained from the IEEE. Contact: Manager, Copyrights and Permissions / IEEE Service Center / 445 Hoes Lane / P.O. Box 1331 / Piscataway, NJ 08855-1331, USA. Telephone: + Intl. 908-562-3966.%
    }
  }

\begin{document}
%
\maketitle
\thispagestyle{firstpage}

\begin{abstract}
Advancements in text-to-image generative AI with large multimodal models are spreading into the field of image compression, creating high-quality representation of images at extremely low bit rates. 
This work introduces novel components to the existing multimodal image semantic compression (MISC) approach, enhancing the quality of the generated images in terms of PSNR and perceptual metrics. The new components include semantic segmentation guidance for the generative decoder, as well as 
content-adaptive diffusion, which controls the number of diffusion steps based on image characteristics.
The results show that our newly introduced methods significantly improve the baseline MISC model while also decreasing the complexity. As a result, both the encoding and decoding time are reduced by more than 36\%. Moreover, the proposed compression framework outperforms mainstream codecs in terms of perceptual similarity and quality. The code and visual examples are available.\footnote{\url{https://github.com/CrashedBboy/SGGIC}}
\end{abstract}

\section{Introduction}
\label{sec:intro}

In 2023, AI-generated images surged to over 15 billion, surpassing traditional stock photo libraries and constituting about a third of all photos ever uploaded to Instagram in just one year~\cite{AIImageStatistics}. Models like DALL-E 2~\cite{ramesh2022hierarchical}, Stable Diffusion~\cite{rombach2022high}, and \href{http://www.overleaf.com}{Midjourney} enable users to create customized, high-quality visuals from simple textual prompts, democratizing creative expression and transforming industries by offering rapid prototyping, concept art, and specialized visuals. This unprecedented scale and flexibility highlight a significant shift in digital content creation, outpacing traditional photography in speed and versatility.

With the exponential increase in images generated from semantic prompts, a shift toward storing images with structural semantics may emerge as a new paradigm in image coding. Semantic-based generative image compression stores images using key descriptive elements like scene and context instead of pixel-level details, enabling higher compression ratios by focusing on semantically relevant aspects. This approach prioritizes semantically significant content and allows flexible editing and dynamic content customization, making it a versatile and efficient paradigm for image storage and manipulation. However, semantic-based image compression faces challenges in maintaining fidelity due to the inherent ambiguity of semantics, as textual prompts often lack detailed elements like textures or lighting, and generative models struggle with precise feature placement~\cite{bordin2023semantic,li2024misc}. Furthermore, the high computational cost of models such as Denoising Diffusion Probabilistic Models (DDPMs)~\cite{ho2020denoising}, which require significant iterative processing, makes them unsuitable for real-time encoding and decoding, posing a significant barrier to practical applications that require speed and efficiency~\cite{ulhaq2022efficient}.

Our work builds upon the multimodal image semantic compression (MISC) approach~\cite{li2024misc}, which is designed to capture essential semantic information from input images to guide a controlled reconstruction process. Our goals are twofold: first, to enhance the reconstruction quality by improving pixel fidelity and maximizing both perceptual similarity and quality; and second, to improve the efficiency of the reconstruction process. To this end, we present the following contributions:
\begin{itemize}
  \item We introduce semantic segmentation \emph{at the decoder} to improve the accuracy of conditional diffusion without the need to encode object masks.
  \item We introduce 
  content-adaptive diffusion to determine the appropriate diffusion settings for each image.
\end{itemize}

\section{Related Work}
In this section, we review related works on generative image codecs that can be broadly classified into Generative Adversarial Network (GAN)-based, Variational Autoencoder (VAE)-based, and diffusion-based approaches. 
Among these, diffusion approaches have gained significant recent interest because of their solid theoretical basis and outstanding performance. 
Their applications in image compression differ in how they employ semantic information. The Sample-What-You-Can-Compress (SWYCC)~\cite{birodkar2024sample} approach combines a fully convolutional autoencoder for representation learning and a diffusion model for image reconstruction and demonstrates significantly better performance compared to GAN-based methods in perceptual fidelity metrics.
Conditional Diffusion Compression (CDC)~\cite{yang2024lossy} encodes a discrete “content” latent variable in the forward diffusion process and uses it for conditioning a denoising diffusion process at the decoder. The method achieves promising results, outperforming GAN baselines in certain metrics. 

Some generative codecs utilize human-readable image semantics. For example, semantic generative compression (SGC)~\cite{bordin2023semantic} and Prompt Inversion Compressor with Sketch (PICS)~\cite{lei2023text+} both utilize latent diffusion models (LDM)~\cite{rombach2022high} to reconstruct images based on various sources of semantic information, including object labels (closed vocabulary), bags of words (open vocabulary), and to map the semantics using a segmentation map or a color map. However, the vagueness introduced by their limited textual guidance results in inconsistency in either color or texture. MISC~\cite{li2024misc} improves the consistency of reconstruction by encoding the image into structural textual semantics and incorporating an autoencoder to provide blurred reconstruction as initial color/pattern guidance for reverse diffusion. However, its patch-based strategy, which uses fixed-size image grid maps to indicate objects’ positions, requires heuristic settings and cannot deal with objects of various sizes. The present work improves upon MISC, enabling better rate-quality trade-offs across various metrics while reducing decoding complexity.

\section{Methods}
\label{sec:methods}
\begin{figure}[tbp]
  \centering
  \includegraphics[width=0.48\textwidth]{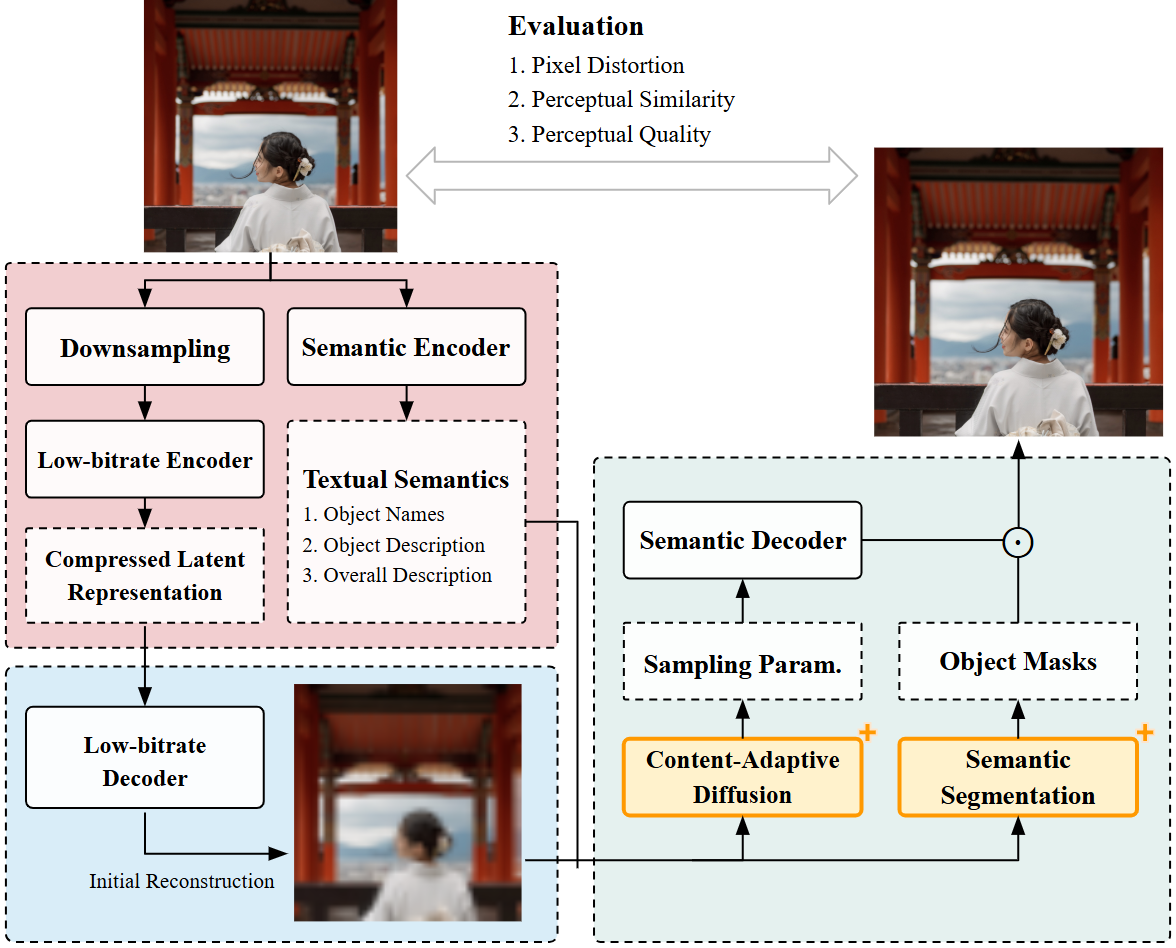}
  \caption{Modules shown in yellow are newly added compared to MISC~\cite{li2024misc}. Content-adaptive diffusion is introduced to adapt diffusion parameters for a particular image. The semantics map encoder of MISC is replaced with the semantic segmentation model ClipSeg~\cite{luddecke2022image} and is relocated to the decoder to reduce bitrate while improving semantic mapping accuracy. 
  }
  \label{fig:pipeline}
\end{figure}

End-to-end compression framework with our proposed methods incorporating with MISC is depicted in Fig.~\ref{fig:pipeline}. The Semantic Encoder extracts textual semantics from the input image, capturing a description of its most prominent objects and an overall scene description. This is accomplished by querying the GPT-4 Vision model~\cite{achiam2023gpt}. These semantics are organized into the set of names $T_n[j]$ and details $T_d[j]$ for the item $j$, and overall description $T_{all}$, together comprising about 60 words. Simultaneously, the input image is downsampled by half and compressed using a low-bitrate autoencoder with the pre-trained model~\cite{cheng2020learned} to produce a compact latent representation, ensuring extreme high compression ratio. Original MISC framework employs a semantic map encoder to code spatial information of various objects in the scene, it is discarded in our encoder design as shown in Fig.~\ref{fig:pipeline}. Instead, such aspects are handled differently using Semantic Segmentation in our proposed decoder.

In the decoding phase, the low-bitrate coded image is first reconstructed with fewer details and blurry textures, which are still sufficient to capture the global structure of the scene. Then, the rest of the decoding processes adds finer details by exploiting the newly introduced Content-Adaptive Diffusion with Semantic Decoder and Semantic Segmentation, eventually improving overall quality of the decoded image. In the remainder of this section, we elaborate further on the decoding process using the two proposed modules.

\subsection{Decoder-side semantic segmentation}
\label{sec:semantic_seg}
Semantic descriptions such as $T_n[j]$, $T_d[j]$, and $T_{all}$ provide a compact scene description of the image but lack spatial information. To compensate for this, MISC encodes spatial information by dividing the image into an $8\times8$ grid and matching patches to textual descriptions using cosine similarity in the CLIP~\cite{CLIP} embedding space. However, this approach requires encoding of patch maps, which increases bitrate, and struggles with precise object boundaries as well as different object sizes, for example very large or very small objects. 

\begin{figure}[tb]
    \centering
    \includegraphics[width=\linewidth]{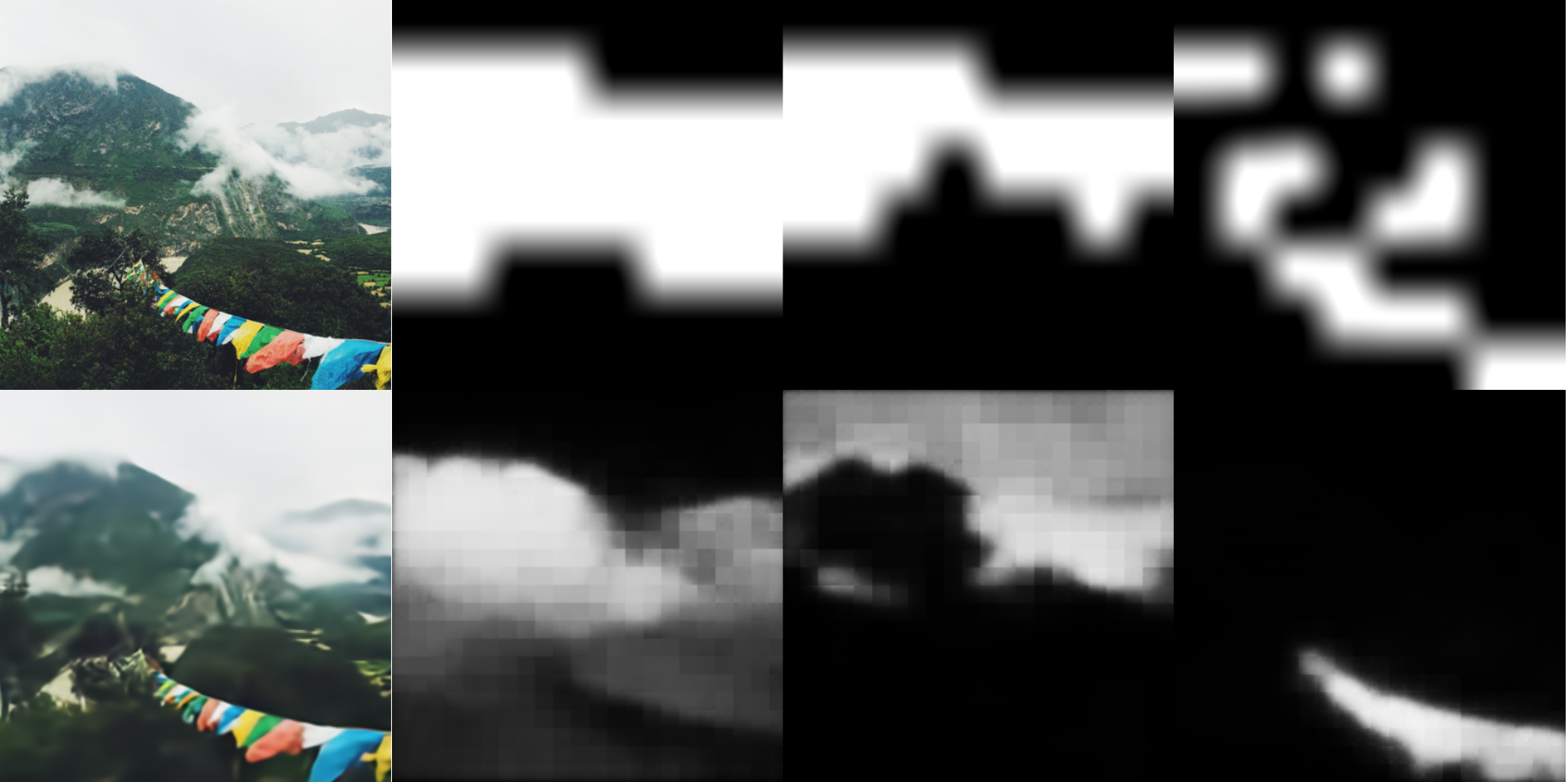}
    \caption{Top, left to right: original image and patch-based semantic maps from MISC for ``Mountain,'' ``Cloud,'' and ``Prayer Flags.'' Bottom, left to right: our decoded low-bitrate image and decoder-side segmentation maps for ``Mountain,'' ``Cloud'' and ``Prayer Flags.''}
    \label{fig:semantic_maps}
\end{figure}

To overcome these limitations, we recognize that a certain amount of \emph{object location information is already available in the low-bitrate image}. Hence, we remove semantic map portion of MISC altogether and avoid coding semantic maps. Instead, we perform semantic segmentation on the low-bitrate image \emph{at the decoder}, based on semantic descriptions. Specifically, we leverage ClipSeg~\cite{luddecke2022image}, a pre-trained open-vocabulary segmentation model with a lightweight conditional decoder to achieve a more precise, adaptable segmentation with zero-shot and one-shot capabilities for object localization based on textual descriptions. A sample result can be seen in Fig.~\ref{fig:semantic_maps}. In the top row, we show the input image and the original MISC semantic maps for ``Mountain,'' ``Cloud,'' and ``Prayer Flags''. In the bottom row, we show the decoded low-bitrate image along with the semantic maps derived from it at the decoder, the newly introduced approach. As seen in the figure, the semantic maps obtained from the decoded low-bitrate image using ClipSeg show more accurate object localization, and they don't require encoding since they can be derived at the decoder.  

\subsection{Content-adaptive diffusion}
\label{sec:diffusion}
Existing diffusion-based image codecs often use fixed diffusion parameters, determined heuristically or through exhaustive testing, which optimize average performance but fail to adapt to individual image characteristics, resulting in suboptimal results for atypical scenes~\cite{li2024misc,birodkar2024sample}. To address this, we introduce content-adaptive diffusion, which adjusts diffusion parameters -- specifically, the number of diffusion sampling steps and the classifier-free guidance ($cfg$) scale -- based on the initial low-bitrate reconstruction and semantics of each image. 

The number of diffusion steps 
significantly impacts reconstructed image appearance and computational cost. Each step refines the previous reconstruction, but the optimal number of steps varies by the image and its content. A higher number of steps enhances details in the reconstruction, but also increases computational cost and may lead to hallucination~\cite{birodkar2024sample}. On the other hand, $cfg$ scale  
controls how strongly the generated output aligns with input guidance, such as textual semantics, via noise prediction: 
\begin{equation}
    \hat{\epsilon}_\theta(x_t | T) = \epsilon_\theta(x_t) + {cfg} \cdot (\epsilon_\theta(x_t | T) - \epsilon_\theta(x_t)),
\label{eq:noise_prediction}
\end{equation}
where \(\epsilon_\theta(x_t | T)\) is the noise estimate conditioned on $T$ (e.g., a text prompt), \(\epsilon_\theta(x_t)\) is the noise estimate without conditioning, and $\hat{\epsilon}_\theta(x_t | T)$ is the updated conditional noise estimate.  
A higher value of \(cfg\) amplifies the influence of conditioning, encouraging 
the model to generate outputs more closely aligned with semantic information $T$. However, due to the usual ambiguity in $T$, parameter $cfg$ should be chosen carefully and according to the image content. 

To predict the 
appropriate number of steps and $cfg$ scale for the particular image, the DynDiff module analyzes the initial low-bitrate reconstructed image and extracts statistical, perceptual, and semantic features. 
Statistical features include edge density and blurriness. Edge density is calculated by converting the grayscale reconstruction to edges using Canny detection~\cite{gonzalez2008digital} and computing the edge-to-pixel ratio. Blurriness is measured using the Laplacian operator~\cite{gonzalez2008digital} on the grayscale image, with its variance representing the level of blurriness. 
Perceptual features are derived from no-reference image quality metrics: BRISQUE~\cite{mittal2011blind}, NIQE~\cite{mittal2012making}, ClipIQA~\cite{wang2023exploring}, MUSIQ~\cite{ke2021musiq}, DBCNN~\cite{networkblind}, and HyperIQA~\cite{su2020blindly}. These metrics provide diverse insights into perceptual quality, leveraging natural scene statistics and multi-scale analysis. 
Semantic features measure the alignment between the initial low-bitrate reconstruction and intended semantics. For this, we use the LIQE metric~\cite{LIQE} and the \textit{semantic 
alignment} $S_a$, calculated as the cosine similarity between CLIP~\cite{CLIP} embeddings of the semantics (\(E_s\)) and the reconstruction (\(E_r\)):
\begin{equation}
    S_a = \frac{E_s \cdot E_r}{\|E_s\| \cdot \|E_r\|}.
    \label{eq:semantic_alignment}
\end{equation}
Together, these features capture a wide range of image characteristics and the alignment of the initial low-bitrate reconstruction with the semantic description. 

\begin{figure*}[t]
\centering
\includegraphics[width=\linewidth]{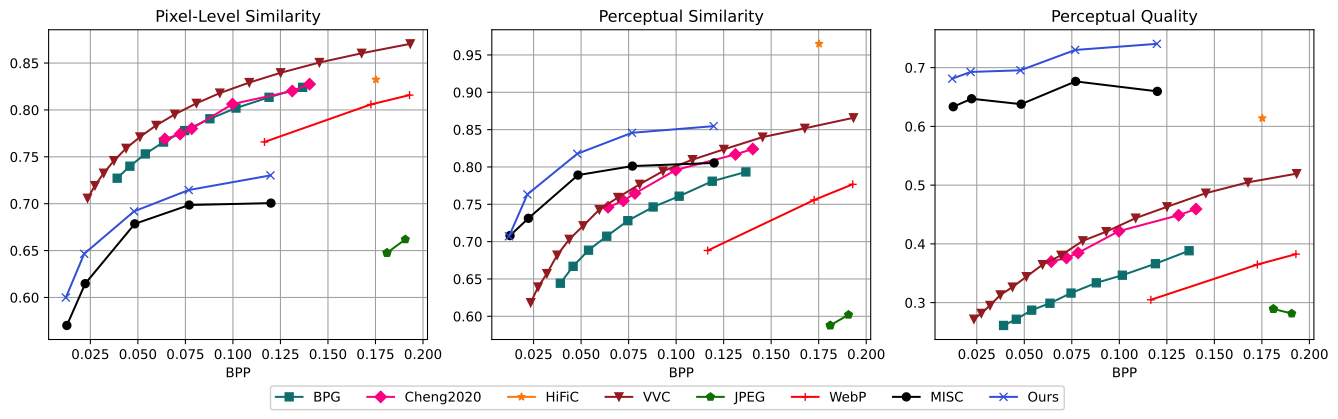}
\caption{Performance of various methods across three groups of image quality metrics: pixel-level similarity, perceptual similarity, and perceptual quality.}
\label{fig:benchmarking}
\end{figure*}

The extracted features are fed to two Multi-Layer Peceptrons (MLPs) to estimate diffusion parameters. One MLP predict the number of diffusion steps, the other predicts the $cfg$ scale. Each MLP has the same architecture -- 3 hidden layers with 16 neurons each -- and produces a scalar in the range $(0,1)$. Finally, the output of the $cfg$ MLP is linearly scaled to $(0,10)$ and the number of steps to $[2,80]$.  
MLPs are trained using the loss function \begin{equation}
    \mathcal{L}_{\text{MLP}} = \frac{1}{n} \sum_{i=1}^n (y_i - \hat{y}_i)^2 + \lambda \cdot \frac{1}{n} \sum_{i=1}^n \hat{y}_i^2,
\label{eq:MLP_loss}
\end{equation}
where $n$ is the batch size, $\hat{y}_i$ is the MLP output for the $i$-th sample in the batch,  $y_i$ is the ground truth that achieves the best LPIPS~\cite{zhang2018perceptual} similarity between the original and reconstructed image, and $\lambda$ is a regularization parameter.  
For the $cfg$ MLP, we set $\lambda=0$ to prioritize perceptual similarity only. For the step prediction MLP, we set $\lambda=0.64$ to encourage a smaller number of steps, thereby reducing the complexity and the chance of hallucination. 

Finally, the semantic decoder utilizes the DiffBIR model \cite{lin2023diffbir}, which integrates ControlNet \cite{zhang2023adding} and latent diffusion models (LDM) \cite{rombach2022high}, to perform conditional diffusion. The reverse diffusion begins with 
the settings predicted by the  
two MLPs described above and the initial low-bitrate reconstructed image. DiffBIR’s blind super-resolution module first upscales the reconstruction by a factor of 2 to match the original input dimensions, expressed as:
\begin{equation}
    x_0 = D_{2\times}(x_{\text{initial}}; T_{\text{all}}),
\label{eq:diffusion_initialization}
\end{equation}
where $D_{2\times}$ is the diffusion process with a $2\times$ upscaling factor, $x_{\text{initial}}$ is the initial reconstruction, and $T_{\text{all}}$ represents all the  semantics used for conditional guidance. The reconstruction is then iteratively refined based on semantics. For each object, its description ($T_d[i]$) and mask ($M[i]$) guide the reverse diffusion, represented as:
\begin{equation}
    x_t = D_{1\times}(x_{t-1}; T_d[i]) \cdot M[i] + x_{t-1} \cdot (1 - M[i]),
\label{eq:diffusion}
\end{equation}
where $x_t$ is the refined reconstruction and $D_{1\times}$ is the diffusion process with 
no upscaling. 
This selective refinement updates only the 
$i$-th object (semantic item), represented by its mask $M[i]$. 
Reconstruction is complete after all objects have been processed.

\section{Experiments and Analysis}

\begin{figure*}[t]
\centering
\includegraphics[width=\linewidth]{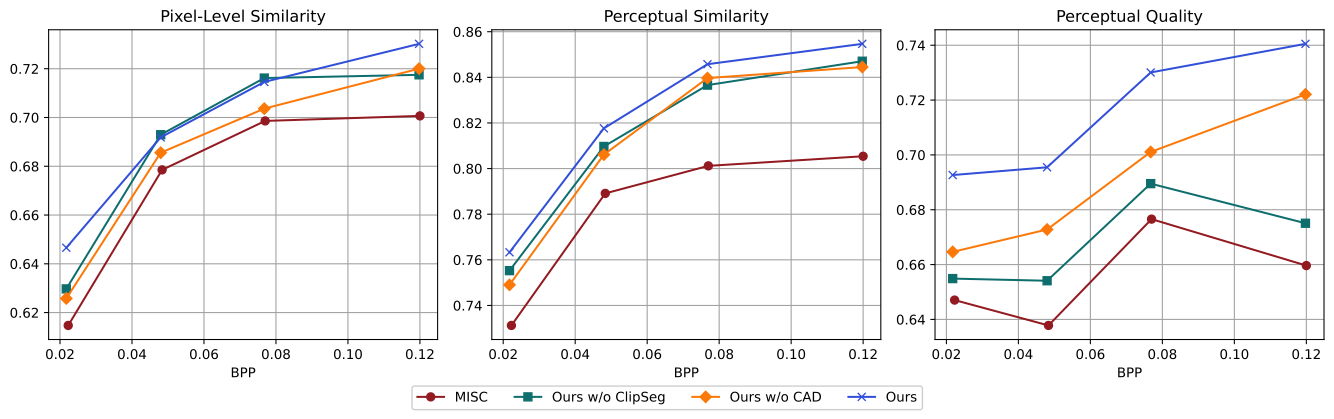}
\caption{Ablation results across various image quality metrics: pixel-level similarity, perceptual similarity, and perceptual quality.}
\label{fig:ablation}
\end{figure*}

\subsection{Settings}

Our method is evaluated on the professional subset of the CLIC2020 dataset~\cite{balle2020nonlinear}, which includes 585 training images and 41 validation images. For consistency, we froze the parameters of the semantics encoder (GPT-4 Vision~\cite{achiam2023gpt}), semantics segmentation model (ClipSeg~\cite{luddecke2022image}), and diffusion models (DiffBIR~\cite{lin2023diffbir}) in the decoder. The training focused on the proposed content-adaptive diffusion and fine-tuning the low-bitrate autoencoder (Cheng2020~\cite{cheng2020learned}).
We benchmarked our method against seven 
image compression methods: traditional codecs (JPEG~\cite{wallace1991jpeg}, WEBP~\cite{WEBP}, BPG~\cite{BPG}, and VVC~\cite{bross2021overview} version VTM 23.6), GAN-based HiFiC~\cite{mentzer2020high}, VAE-based Cheng2020~\cite{cheng2020learned}, and semantic-based MISC~\cite{li2024misc}, upon which our framework was built. 

To assess the quality of reconstructed images, we used three groups of metrics: 
\begin{enumerate}
    \item Pixel-level similarity metrics:  Peak Signal-to-Noise Ratio (PSNR), Structural Similarity Index Metric (SSIM)~\cite{SSIM}, and Multi-Scale SSIM (MS-SSIM)~\cite{MS-SSIM}, which measure pixel-level fidelity relative to the original image;
    \item Perceptual similarity metrics:  CLIPSim~\cite{CLIPSim} and LPIPS~\cite{zhang2018perceptual}, which measure closeness to the original image in learned feature spaces; 
    \item Perceptual quality metrics: ClipIQA~\cite{wang2023exploring}, MUSIQ~\cite{ke2021musiq}, and HyperIQA~\cite{su2020blindly}, which are blind metrics (without reference to the original image) and simply measure the perceptual quality of the output image on its own.
\end{enumerate}

To summarize the performance across these groups of metrics, we 
normalized each metric to [0,1] using IQA-PyTorch-defined\footnote{\url{https://github.com/chaofengc/IQA-PyTorch}} ranges and calculated the arithmetic mean within each group. Subsequently, we show the average performance of each codec across each group of metrics against the bitrate measured in bits per pixel (BPP). 

\subsection{Performance evaluation}

Fig.~\ref{fig:benchmarking} shows the results for the codecs in our experiments.  
The first thing to note is that our approach consistently outperforms MISC across all bitrates and metric groups. This means that the improvements made to the original MISC, described in Section~\ref{sec:methods}, are effective. We dig deeper into each improvement in the ablation study in the next section. 

In terms of comparisons with other codecs, we note that the performance of recent codecs is better than that of MISC or our codec in terms of pixel-level similarity (leftmost graph). This is known from earlier studies, as generative codecs operate with a risk of departure from the original pixel values due to their generative nature. However, it is notable that both MISC and our codec offer better performance at low bitrates than the widely used JPEG.

Where generative codecs shine in comparison with the more conventional codecs is perceptual similarity (middle graph), and especially perceptual quality (rightmost graph), particularly at very low bitrates. Here, their generative nature in combination with compactly encoded semantic information allows producing high-quality images that look perceptually and semantically similar to the original image, although they may lack pixel-value matching fidelity. 

Table~\ref{tab:speed} shows that the performance improvement of our method over MISC did not come at the expense of complexity. The table shows the average encoding and decoding time over the first 20 images of the CLIC2020 validation set measured on a Windows 11 system with a GTX 1060 GPU with 6GB memory, 16GB RAM, Intel i5-7500 CPU, running PyTorch 2.3. Our method saves over 36\% of execution time over MISC in both encoding and decoding. The bulk of the savings in encoding come from avoiding the generation and encoding of the patch-based semantic maps, while savings in the decoding come from the adaptive (and generally lower) number of diffusion steps compared to MISC.

\begin{table}[tb]
    \centering
    \begin{tabular}{l|c|c|c}
        Process & MISC & Ours & \% reduction \\ \hline
         Encoding & 1.19 & 0.76 & 36.13\% \\
         Decoding & 1930.98 & 1194.00 & 38.17\%
    \end{tabular}
    \caption{Average encoding and decoding time in seconds.}
    \label{tab:speed}
\end{table}

\subsection{Ablation study}
Finally, we present the ablation study to gain insight into the contributions brought by the improvements described in Section~\ref{sec:methods}. The baseline is  MISC~\cite{li2024misc} with $cfg=4$ and 40 diffusion steps. We include three versions of our model in the study: the full version (``Ours'' in the figures), an ablated version without ClipSeg, which uses the original patch-based segmentation map coding from MISC (``Ours w/o ClipSeg''), and an ablated version without content-adaptive diffusion (``Ours w/o CAD'').

Fig.~\ref{fig:ablation} shows the results across the three groups of image quality metrics. We note that our full model offers the best results across all cases; its performance is only matched by our model without ClipSeg in terms of pixel-level similarity at bitrates around 0.06 BPP. A possible reason for this is that at these bitrates, the quality of semantic segmentation maps obtained from the initial low-bitrate reconstructed image is not sufficient to provide improvements in terms of pixel-level accuracy, though these maps still provide improvements in the other two groups of metrics.  

Both decoder-side semantic segmentation and content-adaptive diffusion provide clear advantages over the baseline MISC across all metrics. It is interesting to note that decoder-side segmentation is more effective in terms of perceptual quuality improvement (rightmost graph), while content-adaptive diffusion is more effective in improving pixel-level similarity (leftmost graph). Indeed, appropriately adjusting the number of diffusion iterations based on image content reduces the chance of significant departure from the original pixel values due to hallucinations.

\section{Conclusions}

We presented a generative image codec built upon the recent multimodal image semantic compression (MISC) framework, with several important improvements. 
These include the replacement of coarse patch-based semantic maps with decoder-side semantic maps, as well as content-based adaptation of diffusion parameters. 
Extensive experiments were carried out on the CLIC2020 dataset using eight image quality metrics divided into three groups: pixel-level similarity metrics, perceptual similarity metrics, and perceptual quality metrics. The results demonstrated the effectiveness of the newly introduced methods in controlling the diffusion process to produce higher quality images across all metrics and bitrates compared to MISC, while at the same time lowering execution time by over 36\%. In addition, both MISC and the proposed codec showed excellent performance against the more conventional codecs in perceptual similarity and perceptual quality, especially at very low bitrates.

\vfill\pagebreak


\bibliographystyle{IEEEbib}
\small
\bibliography{refs}

\end{document}